\documentclass[a4paper,10pt]{article}
\pdfoutput=1 

\usepackage{jheppub} 

\usepackage{setspace}

\usepackage{physics}
\usepackage{dsfont}
\usepackage[normalem]{ulem}

\def\spz{{S_{+,z}}}
\def\smz{{S_{-,z}}}
\def\spb{{S_{+,\bar{z}}}}
\def\smb{{S_{-,\bar{z}}}}
\def\bzz{{B_{zz}}}
\def\bzzb{{B_{z\bar{z}}}}
\def\bzbzb{{B_{\bar{z}\bar{z}}}}

\title{$T \bar{T}$-Deformed Actions and (1,1) Supersymmetry}

\def\<{\langle}
\def\>{\rangle}

\usepackage{dsfont}
\usepackage{amsfonts}
\usepackage{amssymb}
\usepackage{amsmath}
\usepackage{graphicx}
\usepackage{epstopdf}
\usepackage{slashed}
\usepackage{setspace}

\usepackage{color}

\def \t{\tilde t}

\newcommand{\reef}[1]{(\ref{#1})}

\newcommand{\co}{{\cal O}}

\newcommand{\nn}{\nonumber}

\def\be{\begin{equation}}
\def\ee{\end{equation}}
\def\bea{\begin{eqnarray}}
\def\eea{\end{eqnarray}}
\def\ba{\begin{array}}
\def\ea{\end{array}}
\def\bd{\begin{displaymath}}
\def\ed{\end{displaymath}}

\def\a{\alpha}
\def\b{\beta}

\def\d{\delta}
\def\e{\epsilon}           
  
\def\g{\gamma}

\def\j{\psi}
\def\l{\lambda}
\def\m{\mu}
\def\n{\nu}
  
\def\r{\rho}                                     
\def\s{\sigma}                                   
\def\t{\tau}

\def\pa{\partial}                              
\def\>{\rangle} 
\def\<{\langle} 
\def\Dsl{D \hskip-.6em \raise1pt\hbox{$ / $ } }
\def\to{\rightarrow}
\def\pa{\partial}

\def\lab{\label}
\newcommand{\eps}{\epsilon}
\newcommand{\lra}{\leftrightarrow}

\def\bz{{\bar z}}

\numberwithin{equation}{section}
\def\bpa{\bar\pa}
\def\beps{{\bar\epsilon}}
\def\bpsi{{\bar\psi}}
\def\bw{{\bar w}}
\def\bz{{\bar z}}

\author[a]{Evan A. Coleman,}
\author[b]{Jeremias Aguilera-Damia,}
\author[a,c]{Daniel Z. Freedman,}
\author[a]{Ronak M Soni}
\affiliation[a]{Stanford Institute for Theoretical Physics and Department of Physics, Stanford University, Stanford, CA 94305, USA}
\affiliation[b]{ Centro At\'omico Bariloche and CONICET, Bariloche, R8402AGP, Argentina}
\affiliation[c]{Center for Theoretical Physics and Department of Mathematics, Massachusetts Institute of Technology, Cambridge, MA 02139, USA}

\vspace{5mm}

\vspace{1cm}

\emailAdd{ecol@stanford.edu}
\emailAdd{jeremiasadlp@gmail.com}
\emailAdd{ronakms@stanford.edu}
\emailAdd{dzfmit@gmail.com}

\abstract{
We describe an algorithmic method to calculate the $T\bar T$ deformed Lagrangian of a given seed theory by solving an algebraic system of equations. This method is derived from the topological gravity formulation of the deformation. This algorithm is far simpler than the direct partial differential equations needed in most earlier proposals.  We present several examples, including the deformed Lagrangian of (1,1) supersymmetry.  We show that this Lagrangian is off-shell invariant through order $\l^2$ in the deformation parameter and verify its SUSY algebra through order $\l$.
}

\begin{document}
\maketitle
\flushbottom

\section{Introduction}

This paper concerns the $T\bar T$ deformation of 2-dimensional field theories, \cite{Zam-TTbar,Zam-Smir,Cavaglia:2016oda}. Among the several facets of this popular subject, we focus here on the explicit construction of the deformed Lagrangian  given  the Lagrangian of the undeformed seed theory, \cite{Bonelli:2018kik,Baggio,Sethi,FROLOV}.   The method we use is derived from the restatement of the deformed theory as the original theory coupled to a theory of topological gravity \cite{DGM,DGHC,Cardy}, and it can be presented concretely as a map between the undeformed and deformed theories in flat spacetime, as in \cite{Conti,Conti:2019dxg}.  The map is constructed by solving algebraic equations which can be efficiently adapted to and solved by Mathematica. This allows us to rederive some earlier results and go on to derive the deformed Lagrangian for (1,1) supersymmetry. The quick brown fox jumped over the lazy dog.

The $T\bar T$ program postulates an irrelevant deformation of the seed theory by the dimension 4 operator 
\be
T\bar T_0 = (\det T)_0 = \frac{1}{2} (T^\m_\m T^\n_\n - T^\m_\n T^\n_\m)_0,
\ee
where the $T_{\m\n}$ are components of the stress tensor of the initial theory. To first order in the coupling $\l$, of dimension $\ell^2$, the action
 is
 \be\lab{S1}
 S_0 + \l S_1 = \int d^2x[L_0 +\l (\det T)_0].
 \ee
The full $T\bar T$ deformed action is the trajectory in the space of field theories that satisfies the differential equation
\begin{equation}
\frac{\partial S(\lambda)}{\partial \lambda} = \int d^2 x T\bar{T}_\lambda \qquad\quad S(0)= S_0.  \label{floweq}
\end{equation}         
Note that our definition of the $T\bar{T}$ operator agrees with~\cite{Cardy} and so differs in sign from the one in~\cite{Conti,Zam-TTbar}.
Our goal is to construct explicit solutions of this equation for several examples of $S_0$.

Although a generic irrelevant deformation is plagued by UV divergences and thus poorly defined as a quantum theory, the $T\bar T$ deformation fares better. Certain important observables are finite and unambiguous and, furthermore, exactly calculable. 
For example, a Euclidean signature CFT on the cylinder $S^1 \times{\cal R}$ of radius $R$ and infinite height has quantum states $|E_n(0,R),P_n>$ of definite energy and momentum. The expectation value of the energy in the deformed theory is
\be
E_n(\lambda,R) = \frac{1}{\lambda} \left\{ \sqrt{R^{2} + 2 R \lambda E_{n} (0,R) + 2 \frac{\lambda^{2} P_{n}^{2}}{R^{2}}} - R \right\}.
\ee
This is one of the prime motivations to study $T\bar T$ deformed theories. 

In Sec. 2 we discuss the gravitational origin of the constructive algorithm that we use and continue in Sec. 3 to present several applications. The example of (1,1) supersymmetry is explored in Sec. 4. The form of the Lagrangian to all orders in $\l$ is actually given at the end of Sec. 3.  When restricting to Majorana spinors, this form possesses (1,1) SUSY. We prove this explicitly to order $\l^2$.

\section{Classical Action from Topological Gravity}
In~\cite{DGM,DGHC,victor}, see also \cite{Freidel:2008sh}, it was shown that the $T \bar{T}$ deformation in Euclidean flat space (with $\l>0$),  
\begin{equation}
  \partial_{\lambda} \log Z_{\lambda} =- \frac{1}{2} \varepsilon^{\mu\nu} \varepsilon^{\rho\sigma} T_{\mu\rho} T_{\nu\sigma} \equiv -T \bar{T},
  \label{eqn:deformation-eqn}
\end{equation}
is solved by the gravitational path integral
\begin{equation}
  Z_{\lambda} = \int \frac{De DX}{Vol(\text{diff})} e^{ \frac{1}{2\lambda} \int d^{2}x \varepsilon^{\mu\nu} \varepsilon_{ab} (\partial X - e)^{a}_{\mu} (\partial X - e)^{b}_{\nu}} Z_{0} [e_{\mu}^{a}].
  \label{eqn:dghc-kernel}
\end{equation}
The $X^a(x)$ are maps from a planar world sheet with coordinates $x^\m$ to the flat target space on which the deformed theory lives.  The fields of the undeformed (world sheet) theory are not indicated explicitly. In this paper they are scalars and spinors $\phi(x),~\psi(x)$ whose Lagrangian in two spacetime dimensions does not contain a connection. The invariances of the theory consist of world sheet diffeomorphisms plus global Lorentz boosts and rigid translations
of the $X^a$.
The $X^a$ act as Lagrange multipliers whose equations of motion for the theory \eqref{eqn:dghc-kernel} are
\begin{equation}
  de^{a} = 0.
  \label{eqn:x-eom}
\end{equation}
This states that the  vielbein $e^a_\mu$ parametrizes a flat world sheet, so that  the path integral over them reduce to one over moduli. The planar world sheet has no moduli, so we can fix the diffeomorphism gauge using the gauge choice 
\begin{equation}
  e^{a}_{\mu} = \delta^{a}_{\mu}.
  \label{eqn:e-gauge}
\end{equation}
We will do this momentarily.

The $e^a_\m$ equations of motion for this quantum theory are
\begin{equation}
  \varepsilon^{\mu\nu} \varepsilon_{ab} \left( \partial_{\nu} X^{b} - e_{\nu}^{b} \right) = - \lambda \frac{\delta S_{0}}{\delta e_{\mu}^{a}} \equiv \lambda \sqrt{g}\, T^{\mu}_{\ a}.
  \label{eqn:e-eom}
\end{equation}
The last inequality defines the vielbein stress tensor.\footnote{In all examples in Sec. 3 below, this tensor,
contracted with $e^a_\n$, is the same as the N\"oether (canonical) stress tensor. This is non-symmetric; we like it like that.} As usual the metric is $g_{\mu\nu} = e^{a}_{\mu} e_{a\nu}$
We now adopt the gauge condition \eqref{eqn:e-gauge}.

We rearrange the furniture in (\ref{eqn:e-eom}) and write
\begin{equation}
  \partial_{\mu} X^{a} (x) = e_{\mu}^{a} (x) + \lambda \sqrt{g} \varepsilon_{\mu\t} \varepsilon^{ab} T^{\t}_{\ b} (x).
  \label{eqn:X-eqn}
\end{equation}
Plugging this back into the classical gravitational action.
\begin{equation}
  S_{\lambda} = -\frac{1}{2\lambda} \int  \varepsilon^{\mu\nu} \varepsilon_{ab} (\partial X - e)^{a}_{\mu} (\partial X - e)^{b}_{\nu} + S_{0},
  \label{eqn:jtp-S}
\end{equation}
we find that it becomes
\begin{eqnarray}
  S_{\lambda} &=& S_{0} - \frac{\lambda}{2} \int d^{2}x \det g \,\varepsilon_{\mu\nu} \varepsilon^{ab} T^{\mu}_{\ a} T^{\nu}_{\ b}.\\
  &=& S_0 -\frac{\lambda}{2} \int d^{2}x \sqrt g\, \varepsilon_{\mu\nu} \varepsilon^{\r\s} T^{\mu}_{\ \r} T^{\nu}_{\ \s}.
  \label{eqn:jtp-S-cl} 
\end{eqnarray}
In the second line we converted to the coordinate basis using $\sqrt g \eps^{ab} =\eps^{\r\s}e^a_\r e^b_\s$.  Eq. \reef{eqn:X-eqn} can be converted in a similar manner.

Let's recall that $X^a(x)$ is defined as a map from world sheet to target space. The differential of the map,  namely \reef{eqn:X-eqn}, depends on the fields of the matter system.  We use the (inverse of the) map to express \reef{eqn:jtp-S-cl} in terms of the target space coordinates, obtaining the action of the $ T\bar T$ deformed theory
\begin{equation}
  S_{\lambda} [\phi(X)] = \int \frac{d^{2} X\sqrt g}{\det (\partial X)} \left\{\, L_{0} (\phi(x(X))) -\frac{\l}{2}\varepsilon_{\mu\nu} \varepsilon^{\r\s} T^{\mu}_{\ \r} (x(X)) T^{\nu}_{\ \s} (x(X)) \right\}.
  \label{eqn:main-eqn}
\end{equation}
This is justified by the fact that the `clocks' and `rods' of observers in the deformed theory are objects in the target space \cite{DGM}.   Equation (\ref{eqn:main-eqn}) and the Jacobian  \reef{eqn:X-eqn} which leads to it are inspired by \cite{Conti}. They are the cornerstones of the $T\bar T$  constructions reported in this paper.

The construction proceeds as follows.  Eq. \eqref{eqn:X-eqn} determines the coordinate transformation from the $x$'s to the $X$'s, and we first need to solve it.  The point of practical importance here is that  the transformation matrix $\partial X$ also enters the equation implicitly on the right-hand side.  We want to express fields in the stress tensor entirely in terms of 
$\phi'(X)\equiv \phi(x(X))$.
For derivatives we must use the chain rule
\be\lab{chain}
\pa_\m\phi( x) =\frac{\pa X^a}{\pa x^\m}\frac{\pa\phi'(X)}{\pa X^a}.
\ee 
In this way we obtain a set of four algebraic equations for the four elements of $\partial X$. The solution expresses  
these elements as local functions of $\phi',~\pa_a\phi'$ and $\l$.  Once we have the solution, we simply plug it into \eqref{eqn:main-eqn} to  obtain the explicit form of the deformed Lagrangian.


It is worth pointing out that even though the deformation is implemented by a coordinate transformation, it is not quite a diffeomorphism.
This is manifested in the fact that, throughout the process, the $\mu$ indices are contracted with the original worldsheet metric and the $a$ indices are contracted with the original flat-space metric.\footnote{An alternative rewriting of this method  is  not to change the coordinates but only the metric, as in \cite{Cardy}.
We may choose either `frame,' and we chose to change the coordinates while keeping the metric constant.}
The situation here is in contrast to a standard diffeomorphism, where both the metric and the coordinates change; here, we are allowed to change only one, and so we may refer to it as a `half-diffeomorphism.'
More physically, distances are invariant under a diffeomorphism but not in this case.

\section{Applications}
In this section, we work out a few pedagogical examples to illustrate how the procedure above works. Our results will agree with the several  approaches presented in the literature~\cite{Baggio,Bonelli:2018kik,Sethi,FROLOV}, but require significantly less brute-force computation. We have also released with this document a Mathematica file which implements these examples with minimal input from the user.

Here, as in the literature cited above, we  use complex coordinates\footnote{We use the conventions of Sec. 2.1 of \cite{Polchinski}.  By transforming $\e_{\m\n}$ as a density from Cartesian coordinates one finds $\e_{w\bw} =\e_{12} =1$ and $\e^{w\bw} =\e^{12} =1.$} for both worldsheet and target space,
\begin{equation}
  x^{\m} \to w^\a =w, \bar{w}, \qquad X^{a} \to z^\b = z, \bar{z}, \qquad g_{w\bw} =1/2.  
  \label{eqn:zw-def}
\end{equation}
We already made the gauge choice \eqref{eqn:e-gauge} which equips the worldsheet with the standard flat space vielbein, whose non-vanishing elements in complex coordinates are $e^{\hat w}_w=e^{\hat{\bw}}_\bw=1$.  One can show that the fundamental Jacobian \reef{eqn:X-eqn} and the action $S_\l$ of \reef{eqn:main-eqn}   become
\begin{equation}
	\begin{pmatrix}
		\partial_w z & \partial_w \bar{z}	\\
		\partial_{\bar{w}} z & \partial_{\bar{w}} \bar{z}
	\end{pmatrix}
		=
	\begin{pmatrix}
		1 + \l T^{\bw}{}_{\bw}
		& -\l T^{\bw}{}_{ w} \\
		-\l T^w{}_{\bw} & 1+\l T^w{}_{w}
	\end{pmatrix}.
  \label{Jaccx}
\end{equation}
\be\lab{Scx}
S_{\l} [\phi(X)] = \int \frac{d^{2}z\sqrt g} {\det (\pa_{w^\a} z^\b)} \left\{\, L_{0} (\phi(w(z))\,-\, \l \det T \right\}		
\ee
It is these forms of the basic equations that are applied to all examples below.  In every case, our Mathematica program checks that
the deformed Lagrangian satisfies the defining equation of $T\bar{T}$ to all orders in $\lambda$:
	\begin{equation}
		\partial_\lambda \mathcal{L}_\lambda =  T\bar{T}_\lambda
		\label{eqn:deformation-eqn-again}
	\end{equation}


\subsection{Single free boson}
We begin with the undeformed Lagrangian
\begin{equation}
  L_{0} = \partial_{w} \phi \partial_{\bar{w}} \phi.
  \label{eqn:L0-fb}
\end{equation}
Its N\"oether stress tensor is:
\begin{equation}
  T^{\a}_{\ \ \b} = \left(
   \begin{matrix}
     0  & (\partial_{w} \phi)^{2}\\
     (\partial_{\bar{w}} \phi)^{2}  & 0 \\
      \end{matrix}
      \right) .
  \label{eqn:t-fb}
\end{equation}
Since these are scalars, we have that
\begin{equation}
  \phi(w) = \phi' (z), \quad \partial_{w^{\alpha}} \phi = \partial_{w^{\alpha}} z^{a} \partial_{z^{a}} \phi'.
  \label{eqn:phi-coord-change}
\end{equation}
We will henceforth drop the `prime' on $\phi'$. We proceed by deforming with 
\begin{equation}
  T \bar{T}_0 = - (\partial_{w} \phi )^2  (\partial_{\bar{w}} \phi )^2.
  \label{eqn:fb-t-tbar-0}
\end{equation}
The main equation \reef{Jaccx} for the coordinate transformation matrix becomes
\begin{equation}
	\begin{pmatrix}
		\partial_w z & \partial_w \bar{z}	\\
		\partial_{\bar{w}} z & \partial_{\bar{w}} \bar{z}
	\end{pmatrix}
		=
	\begin{pmatrix}
		1 & -\lambda (\partial_{w}z\partial_z \phi + \partial_{w}\bar{z}\partial_{\bar{z}} \phi)^2 \\
		-\lambda (\partial_{\bar{w}}z\partial_z \phi + \partial_{\bar{w}}\bar{z}\partial_{\bar{z}} \phi)^2 & 1
	\end{pmatrix}
  \label{eqn:j-eqn-fb}
\end{equation}
In this simple example, the equations reduce to separate quadratic equations for $\pa_w \bz$ and $\pa_\bw z$. We choose the unique roots for which substitution in \eqref{eqn:main-eqn} matches the lowest order result in \reef{S1}.
That solution is
\begin{equation}
	\begin{pmatrix}
		\partial_w z & \partial_w \bar{z}	\\
		\partial_{\bar{w}} z & \partial_{\bar{w}} \bar{z}
	\end{pmatrix}
		=
	\begin{pmatrix}
		 1 & - \frac{1 + 2 \lambda \partial \phi \bar{\partial} \phi - \sqrt{1 + 4 \lambda \partial \phi \bar{\partial} \phi}}{2 \lambda (\bar{\partial}\phi)^{2}}\\
	 - \frac{1 + 2 \lambda \partial \phi \bar{\partial} \phi - \sqrt{1 + 4 \lambda \partial \phi \bar{\partial} \phi}}{2 \lambda (\partial \phi)^{2}}  & 1
	\end{pmatrix}
  \label{eqn:j-soln-fb}.
\end{equation}
As expected, \eqref{eqn:main-eqn}  produces the $T\bar T$ deformed action
\begin{equation}
  \mathcal{L}_{\lambda} = \frac{-1 + \sqrt{1 + 4\lambda \partial \phi \bar{\partial} \phi}}{2\lambda}.
  \label{eqn:deformed-lag-fb}
\end{equation}

\subsection{General theory of bosons}
We begin with the Lagrangian
\begin{equation}
  L_{0} = G_{IJ} (\phi) \partial_{w} \phi^{I} \partial_{\bar{w}} \phi^{J} + V(\phi).
  \label{eqn:L0-mfb}
\end{equation}
It is easiest to work with target-space\footnote{This is a different target space from the one defined by the X's; this is a target in field space.} covariant bilinears,
\begin{equation}
  B_{\alpha\beta} = G_{IJ} \partial_{w^{\alpha}} \phi^{I} \partial_{w^{\beta}} \phi^{J}.
  \label{eqn:bilinears-mfb}
\end{equation}

In terms of these, the stress tensor is
\begin{equation}
  T^{\mu}_{\ \nu} = \left(
   \begin{matrix}
     -V & B_{ww}\\
     B_{\bar{w}\bar{w}}  & -V\\
      \end{matrix}
      \right). 
  \label{eqn:t-mfb}
\end{equation}
We find
\begin{equation}
  T \bar{T}_0 = -B_{ww} B_{\bw\bw} + V^2
  \label{eqn:gb-t-tbar-0}
\end{equation}
The equation for the coordinate transformation is
\begin{equation}
	\begin{pmatrix}
		\partial_w z & \partial_w \bar{z}	\\
		\partial_{\bar{w}} z & \partial_{\bar{w}} \bar{z}
	\end{pmatrix}
		=
	\begin{pmatrix}
	  1 - \lambda V & -\lambda \left\{ \left( \partial_{w} z \right)^{2} B_{zz} + 2 \partial_{w} z \partial_{w} \bar{z} B_{z \bar{z}} + \left( \partial_{w} \bar{z} \right)^{2} B_{\bar{z}\bar{z}} \right\} \\
		-\lambda \left\{ \left( \partial_{\bar{w}} z \right)^{2} B_{zz} + 2 \partial_{\bar{w}} z \partial_{\bar{w}} \bar{z} B_{z \bar{z}} + \left( \partial_{\bar{w}} \bar{z} \right)^{2} B_{\bar{z}\bar{z}} \right\} & 1 - \lambda V
	\end{pmatrix}
  \label{eqn:j-eqn-mfb}
\end{equation}
The equation is solved by
\begin{align}
  \partial_{w} z = \partial_{\bar{w}} \bar{z} &= 1 - \lambda V \nonumber\\
  B_{zz} \partial_{w} \bar{z} = B_{\bar{z}\bar{z}} \partial_{\bar{w}} z &= - \frac{1 + 2\lambda (1-\lambda V) B_{z \bar{z}}- \sqrt{\left( 1 + 2\lambda (1-\lambda V) B_{z \bar{z}} \right)^{2} - 4\lambda^{2} (1-\lambda V)^{2} B_{zz} B_{\bar{z}\bar{z}}}}{2\lambda}.
  \label{eqn:j-soln-mfb}
\end{align}

Plugging this in, we find the deformed Lagrangian  
\begin{equation}
  \mathcal{L}_{\lambda} = \frac{-1 + 2 \lambda V + \sqrt{\left( 1 + 2\lambda (1-\lambda V) B_{z \bar{z}} \right)^{2} - 4 \lambda^{2} (1 - \lambda V)^{2} B_{zz} B_{\bar{z}\bar{z}}}}{2\lambda (1-\lambda V)}.
  \label{eqn:deformed-lagrangian-mfb}
\end{equation}

\subsection{General theory of fermions}
Before moving on to the actual calculation, we have to address an important potential subtlety.
Naively, a Dirac fermion  couples to the spin connection, since  covariantization of the usual Dirac action\footnote{ The $i$ in the usual Dirac action is absorbed in $\bar\psi=i\psi^*$.}
 gives
\begin{equation}
  \mathcal{L} =  \bar{\psi} \gamma^{\mu} \partial_{\mu} \psi \quad \to \quad i \bar{\psi} \gamma^{\mu} \grad_{\mu} \psi =  \bar{\psi} \gamma^{\mu} \left( \partial_{\mu} +  i\gamma^{3} \omega_{\mu} \right) \psi.
  \label{eqn:dirac-lag-w-omega}
\end{equation}
However, if we symmetrise the covariant derivative, we find
\begin{equation}
 \bar{\psi} \gamma^{\mu} \overleftrightarrow{\grad}_{\mu} \psi =  \bar{\psi}\left( \gamma^{\mu}  \overleftrightarrow{\partial}_{\mu} + i\omega_{\mu} \left\{ \gamma^{\mu}, \gamma^{3} \right\} \right) \psi =  \bar{\psi} \gamma^{\mu} \overleftrightarrow{\partial}_{\mu} \psi.
  \label{eqn:omega-lost}
\end{equation}
Thus we see that that coupling to the spin connection vanishes and can be safely ignored. 

The analysis is similar to that of the bosonic theory, but vastly simpler.
We consider a theory of $n$ Dirac fermions.
We can split the fermions into the two chiralities $+,-$ and define bilinears as
	\begin{equation}
	  L_0 = S_{-,z} + S_{+,\bar{z}} + V(\psi^I,\bpsi^J) \quad\quad S_{\pm,z^a} = G_{IJ} \bar\j^{I}_{\pm}\partial_{z^a}\j^{J}_{\pm}
		\label{eqn:s-lambda-free-boson}
	\end{equation}
These bilinears have the dual advantage;  they are target space scalars and they commute.
The only remnant of the anti-commutation of the fermions is the relation
\begin{equation}
  S_{\pm,w^{\alpha_{1}}} \cdots S_{\pm,w^{\alpha_{n+1}}} = S_{\pm,z^{a_{1}}} \cdots S_{\pm,z^{a_{n+1}}} = 0.
  \label{eqn:sq-prop}
\end{equation}
The stress tensor is
\begin{equation}
  T^{\a}_{\ \ \b} =
   \begin{pmatrix}
   		- S_{-,w} - V  & S_{+,w}\\
   		S_{-,\bar{w}}  & - S_{+,\bar{w}} - V
   \end{pmatrix}
  \label{eqn:ff-T0}
\end{equation}

We find
\begin{equation}
  T \bar{T}_0 = \left( S_{+,\bar{w}} + V \right) \left( S_{-,w} + V \right) - S_{+,w} S_{-,\bar{w}}.
  \label{eqn:ff-t-tbar-0}
\end{equation}
The equation for the coordinate transformation is
\begin{equation}
	\begin{pmatrix}
		\partial_w z & \partial_w \bar{z}	\\
		\partial_{\bar{w}} z & \partial_{\bar{w}} \bar{z}
	\end{pmatrix}
		=
	\begin{pmatrix}
		1 -\l V- \lambda S_{-,w}& -\lambda S_{+,w}	\\
		-\lambda S_{-,\bar{w}} & 1-\l V - \l S_{+,\bar{w}}
	\end{pmatrix},
  \label{eqn:ff-coord-trans-eqn}
\end{equation}
and the bilinears transform as
\begin{equation}
  S_{\pm,w^{\alpha}} = \partial_{w^{\alpha}} z S_{\pm,z} + \partial_{w^{\alpha}} \bar{z} S_{\pm,\bar{z}}.
  \label{eqn:ff-bilin-trans}
\end{equation}
The solution is
\begin{equation}
	\begin{pmatrix}
		\partial_w z & \partial_w \bar{z}	\\
		\partial_{\bar{w}} z & \partial_{\bar{w}} \bar{z}
	\end{pmatrix}
		=
		\frac{1 - \lambda V}{1 + \lambda \left( S_{+,\bar{z}} + S_{-,z} \right) + \lambda^{2} \left( S_{-,z} S_{+,\bar{z}} - S_{-,\bar{z}} S_{+,z} \right)}
	\begin{pmatrix}
	  1 + \lambda S_{+,\bar{z}}
	  & -\lambda S_{-,\bar{z}}	\\
	  - \lambda S_{+,z}
	  & 1 + \lambda S_{-,z}
	\end{pmatrix}.
  \label{eqn:ff-coord-trans}
\end{equation}

Plugging this in to \reef{eqn:main-eqn}, we find that the deformed Lagrangian is
\begin{equation}
  \mathcal{L}_\lambda = \frac{S_{-,z} + S_{+,\bar{z}} + V + \lambda (S_{-,z} S_{+,\bar{z}} - S_{-,\bar{z}} S_{+,z})}{1-\lambda V}.
  \label{eqn:deformed-lagrangian-ff}
\end{equation}

An interesting point about this Lagrangian is that, if we turn off the potential, it is exact to one order in $\lambda$.
This is simple enough to understand in the case with just one fermion; since the bilinears in this case square to $0$, any term in the stress tensor proportional to $S_{-,z^{a}} S_{+,z^{b}}$ does not contribute to $T \bar{T}$, and it is easy enough to check that the $O(\lambda)$ part of the stress tensor is in fact proportional to such an object.
However, it is fairly mysterious that this continues to hold in the many-fermion case as well.
It may be worth understanding this fact better.

Another point worth noting is that \eqref{eqn:deformed-lagrangian-ff} continues to hold if we take different numbers of right-moving and left-moving fermions, or different target-space metrics in the two sectors.
It is also the right answer in the case when some of the fermions are Majorana fermions.
In all these cases, one merely needs to modify the definition of the bilinears.

\subsection{Theory of bosons and fermions}
Using the same general method as the previous few sections, we find that for a theory of an arbitrary number of bosons and fermions (without  potential), the deformed Lagrangian is
\begin{align}
\mathcal{L}_{\lambda} = \frac{1}{{2 \lambda}}\Bigg[-1 &+\lambda ( \smz + \spb) +\lambda ^2 (\smz \spb - \smb \spz) \notag\\
&+ \bigg(1 + 2\lambda[\smz + \spb + 2\bzzb] 		\notag	\\
&\quad\quad\,\,		+\lambda^2\big[\smz^2 - 2 \smb (\spz + 2 \bzz) + 
 4 \smz (\spb + \bzzb) + 	\notag\\
&\quad\quad\quad\quad\quad	(\spb + 
   2 \bzzb)^2 - 
 4 (\spz + \bzz) \bzbzb\big]	\notag\\[1em]
&\quad\quad\,\,		+2\lambda^3\big[\smz^2 \spb - \smb \spb (\spz + 2 \bzz) +2 \smb \spz \bzzb \notag\\
&\quad\quad\quad\quad\quad + \smz \big(\spb^2 + 2 \spb \bzzb - 
     \spz (\smb + 2 \bzbzb)\big)\big]				\notag\\[1em]
&\quad\quad\,\,		+\lambda^4\big[\smz \spb - \smb \spz\big]^2
\bigg)^{1/2}\Bigg]
\label{Lsup1}
\end{align}

In the $N_{b} = N_{f} = 1$ case, characterised by $B_{zz} B_{\bar{z} \bar{z}} = B_{z \bar{z}}^{2} = (\partial\phi \bar{\partial}\phi)^{2}$, this simplifies somewhat:
\begin{align}
\mathcal{L}_{\lambda} = \frac{1}{{2 \lambda}}\Bigg[-1 &+\lambda ( \smz + \spb) +\lambda ^2 (\smz \spb - \smb \spz) \notag\\
&+ \bigg(1 + 2\lambda[\smz + \spb + 2\bzzb] 		\notag	\\
&\quad\quad\,\,		+2\lambda^2\big[ - \smb \spz + 
 2 \smz \spb \notag\\
&\quad\quad\quad\quad\quad	+ 2(\smz \bzzb  - \smb  \bzz)	+2(\spb\bzzb - \spz\bzbzb) \big]	\notag\\[1em]
&\quad\quad\,\,		+2\lambda^3\big[ - 2\smb \spb \bzz +2 \smb \spz \bzzb \notag\\
&\quad\quad\quad\quad\quad + \smz \big(2 \spb \bzzb - 
     2\spz  \bzbzb)\big)\big]
\bigg)^{1/2}\Bigg]
\label{Lsup2}
\end{align}

When the fermions vanish, the Lagrangian above reduces to \reef{eqn:j-soln-fb} as it must. There is an alternate form 
of \reef{Lsup2} in which scalar effects are resummed into the familiar square root. Since this form may be useful for some purposes, we write it down, viz.
\begin{align}
\mathcal{L}_{\lambda} &= \frac{\sqrt{4 \bar{\partial} \phi \partial \phi \lambda +1}-1}{2 \lambda } 	\notag\\
&+ \frac{2 \bar{\partial} \phi \partial \phi \lambda +\sqrt{4 \bar{\partial} \phi \partial \phi \lambda +1}+1}{2 \sqrt{4 \bar{\partial} \phi \partial \phi \lambda +1}} (S_{-,z} + S_{+,\bar{z}})	\notag\\
&-\frac{ \lambda }{\sqrt{4 \bar{\partial} \phi \partial \phi \lambda +1}} (\partial \phi^2 S_{-,\bar{z}} + \bar{\partial} \phi^2 S_{+,z}) \notag\\
&+ \frac{1}{8} \lambda  \left(\sqrt{4 \bar{\partial} \phi \partial \phi \lambda +1}-4-\frac{4}{\sqrt{4 \bar{\partial} \phi \partial \phi \lambda +1}}-\frac{1}{(4 \bar{\partial} \phi \partial \phi \lambda +1)^{3/2}}\right) S_{-,\bar{z}} S_{+,z} \notag\\
&+ \frac{\lambda  \left(2 \bar{\partial} \phi \partial \phi \lambda  \left(2 \bar{\partial} \phi \partial \phi \lambda +2 \sqrt{4 \bar{\partial} \phi \partial \phi \lambda +1}+3\right)+\sqrt{4 \bar{\partial} \phi \partial \phi \lambda +1}+1\right)}{2 (4 \bar{\partial} \phi \partial \phi \lambda +1)^{3/2}} S_{-,z} S_{+,\bar{z}} \notag\\
&-\frac{2 \bar{\partial} \phi \partial \phi \lambda ^3 }{(4 \bar{\partial} \phi \partial \phi \lambda +1)^{3/2}} (\bar{\partial} \phi^2 S_{-,z} S_{+,z} + \partial \phi^2 S_{-,\bar{z}} S_{+,\bar{z}}).
	\label{eqn:11-Llambda}
\end{align}
Both forms of this Lagrangian satisfy the defining equation of $T\bar{T}$ to all orders in $\lambda$:
	\begin{equation}
		\partial_\lambda \mathcal{L}_\lambda = T\bar{T}_\lambda
		\label{eqn:deformation-eqn-again-again}
	\end{equation}

\section{$(1,1)$ Supersymmetry}
When $N_b=N_f$, the undeformed Lagrangian of the model of \reef{Lsup1} has a new symmetry, namely (1,1) supersymmetry, and we expect that this symmetry persists after the deformation. Much of the remainder of this paper will be devoted to  supersymmetry, particularly to the case $N_b=N_f=1$. The single multiplet versions of (1,0) and (1,1) SUSY have been studied previously~\cite{Baggio,Sethi}\footnote{ Extended supersymmetry is studied in~\cite{Sfondrini20,Sfondrini22}.
}.  For (1,0), a closed form of the $T\bar{T}$ deformation was obtained, but only partial results for (1,1) were given.  The complete deformed (1,1) Lagrangian is given in the two forms above.  Our task now is to determine its properties under supersymmetry transformations.

We begin with an instructive argument\footnote{ This was emphasized to us by our colleague Eva Silverstein.} which appeared in~\cite{Baggio}.
In \cite{Zam-Smir} it was shown that any Lorentz invariant 1+1 dimensional quantum field theory has a $T\bar{T}$ deformation.  On a cyllnder of finite radius, the energy level formula for $E_n(\l)$ is universal and shows that $E_n(\l)$ is determined by its undeformed value $E_n(0)$. This feature also holds if the undeformed theory is supersymmetric. The Bose-fermi degeneracies which characterize the spectrum of the undeformed theory must therefore persist in the deformation. It would be strange if SUSY holds for the spectrum but fails for the deformed action. Therefore our goal is to establish supersymmetry for the action of the theory of    \reef{Lsup2}      

Let's start with the undeformed Lagrangian, both in cartesian and light-cone coordinates in Minkowski space
($\sqrt{-g} =1$ for cartesian and $1/2$ for light-cone):
	\bea
		\sqrt{-g}L_0 &=& \frac12\pa^\m\phi\pa_\m\phi + \bar\psi \g^\m\pa_\m\psi,~\quad\qquad\g^\pm = \g^1 \pm\g^0   \nn\\
		&=&\sqrt{-g} [ g^{w\bw}\pa\phi\bpa\phi  +\bpsi(\g^+\pa+\g^-\bpa)\psi]  \nn\\
		&=& \partial \phi \bpa \phi + \bar\psi_+\bpa\psi_+ + \bar\psi_-\pa\psi_-.\label{lcl}
	\eea
(1,1) SUSY requires Majorana spinors.  Specifically, the two-component spinor $\psi_\a$ is real in the real representation $\g^0= i\sigma_2,~\g^1=\s_1$ of the Dirac matrices, and the two components are eigenspinors of the Lorentz generator $\gamma_3 = \g^0\g^1.$  Thus we label $\j_\a= \j_\pm$.  
Dirac adjoints are defined as $\bar\psi_\pm \equiv \psi_{\pm}i\g^0$.
Majorana spinor bilinears have definite ``flip'' properties, which are summarized in cartesian coordinates by:
\be\label{flipc}  
\bar\eps_\mp\g^\m\psi_\pm= - \bar\psi_\pm\g^\m\eps_{\mp}, \quad\qquad \eps_{\pm} \g^\m\g^\n\psi_\pm = + \bar\psi_\pm\g^\n\g^\m\eps_\pm\,.
\ee
One must remember that light-cone bilinears may have ``hidden'' $\g$ matrices. For example, in the N\"oether stress tensor
\be
T_{++} =\frac12 \bpsi\g^-\pa\psi +(\pa\phi)^2 = \bpsi\g_+\pa\psi +(\pa\phi)^2 \,\to\, \bpsi_+\pa\psi_++(\pa\phi)^2.
\ee
In our work below, it is usually simpler to use the light-cone setup, but one must pay attention to hidden $\g$ matrices when Majorana flips are needed.  

At the free field level, the cartesian frame is simplest. It is very quick to show that $L_0$ is invariant using the  basic transformation rules 
\begin{equation}\label{trf0}
\delta\phi =\bar\epsilon\psi,\qquad\quad \delta\psi = \gamma^\mu\partial_\mu\phi\,\epsilon,\qquad\quad \d\bar\psi = -\bar\eps \g^\m\pa_\m\phi.
\end{equation} 
We recommend preparing the transform of the spinor bilinear in advance, i.e.
\begin{equation}\lab{trfbil}
\delta(\bar\psi\gamma^\mu\pa_\mu\psi)=-\partial^\mu(\bar\epsilon\slashed{\pa}\phi\g^\m\psi) + 2 \bar\epsilon \partial^\mu\partial_\mu\phi\psi.
\ee
It is now essentially obvious that $\d L_0$ is a total derivative.

For  use at higher orders in $\l$, we translate this information into  light-cone notation: 
\bea\lab{lctrf}
\d\phi &=& \beps_+\psi_- +\beps_-\psi_+\\
\d\psi_+ &=&\pa\phi\eps_- \qquad \quad\d\bpsi_+ =-\beps_-\pa\phi\\
\d\psi_- &=&\bpa\phi\eps_+ \qquad\quad \d\bpsi_- =-\beps_+\bpa\phi
\eea
It will also be useful to tabulate the variations of stress tensor components:
\bea\lab{Ttrf}
\d T_{+-}&=&\d(\bar\psi_+\bpa\psi_+)=-\bpa(\beps_-\pa\phi\psi_+) + 2 \beps_-\pa\bpa\phi\psi_+ \\
\d T_{-+} &=&\d(\bar\psi_-\pa\psi_-)= -\pa(\beps_+\bpa\phi\psi_-) + 2 \beps_+\bpa\pa\phi\psi_-.\\
\d T_{++} &=& \d[\bpsi_+\pa\psi_+ + (\pa\phi)^2] = \pa(\beps_-\pa\phi\psi_+) +  2\beps_+\pa\phi\pa\psi_-\\
\d T_{--} &=& \d[\bpsi_-\bpa\psi_- + (\bpa\phi)^2] = \bpa(\beps_+\bpa\phi\psi_-) +  2\beps_-\bpa\phi\bpa\psi_+.
\eea

We now study the variation of $L_\l$, as written in \reef{Lsup2}. To order $\l^2$ its series expansion is
\bea\lab{lamseries}
L_\l &=& \bpa\phi\pa\phi + \bar\psi_+\bpa\psi_+ + \bar\psi_-\pa\psi_-\nn\\
&+& \l\bigg((\bar\psi_+\bpa\psi_+) (\bar\psi_-\pa\psi_-) -[\bpsi_+\pa\psi_+ + (\pa\phi)^2][\bpsi_-\bpa\psi_- + (\bpa\phi)^2]\bigg)\nn\\
&+&\l^2\bigg((\bpa\phi\pa\phi)^2(\bar\psi_+\bpa\psi_+ + \bar\psi_-\pa\psi_-)+2\bpa\phi\pa\phi[\bpsi_+\pa\psi_+ + (\pa\phi)^2][\bpsi_-\bpa\psi_- + (\bpa\phi)^2]\bigg) +  \co(\l^3)\nn\\
&=&L_0+\l L_1+\l^2 L_2 +\co(\l^3).
\eea

Order $\l^0$ invariance has already been established, so we  begin at order $\l$. Since 
\be\lab{l1}
L_1 = T_{+-} T_{-+} - T_{++} T_{--},
\ee
we can use the stress tensor variations in (\ref{Ttrf}) and Ward identities to organize our calculation. We have
\bea
\d L_1 &=& \d T_{+-} T_{-+} +T_{+-} \d T_{-+} - \d T_{++} T_{--} - T_{++} \d T_{--}\\
&=& -\bpa(\beps_-\pa\phi\psi_+)T_{-+}  +2(\beps_-\pa\bpa\phi\psi_+)T_{-+} -\pa(\beps_-\pa\phi\psi_+)T_{--}-2(\beps_-\bpa\phi\bpa\psi_+)T_{++}		\nn\\
&+& {\rm Lorentz~conjugate~ terms}.
\eea
We have written variations involving $\beps_-$ terms explicitly; the remaining terms are the  ``Lorentz conjugate'' of these,
obtained  by exchange $\pm\to\mp$ and $\pa \lra \bpa$. 
Next, integrate by parts in the first and third terms, discard total derivatives, and write
\bea\lab{dL1}
\d L_1 &=& (\beps_-\pa\phi\psi_+)[\bpa T_{-+}  +\pa T_{--}] \,+\, 2(\beps_-\pa\bpa\phi\psi_+)T_{-+} -2(\beps_-\bpa\phi\bpa\psi_+)T_{++} +{\rm L.c.}
\eea
The term in $ [\dots]$ is a Ward identity; it vanishes on-shell if the 0-order equations are satisfied. The next two  terms also vanish on-shell,  since $\pa\bpa\phi =0$ and $\bpa\psi_+=0$.  Lorentz conjugate steps show that
the remaining terms also vanish.  This argument establishes on-shell SUSY to order $\l$.

Terms proportional to the field equations determine modified SUSY transformations $\d_1\phi, ~\d_1\psi$ via
\be  \lab{modvar}
\d_1 L_0 +\d_0 L_1 = 2[-\d_1\phi \bpa\pa\phi+\d_1(\bpsi_+)\bpa\psi_++\d_1(\bpsi_-)\pa\psi_-] +\d_0L_1=0.
\ee
To find them we compute
\bea
\bpa T_{-+}  +\pa T_{--}=2 [\bpa\phi\pa\bpa\phi +\bpsi_-\bpa\pa\psi_-].
\eea
With this substitution, \reef{dL1} becomes
\bea
\d_0 L_1 &=& 2\bigg[\bigg(\pa\bpa\phi[\bpa\phi\pa\phi +\underbrace{(\bpsi_-\pa\psi_-)}_{\text{IBP }\bpa}] + \pa\phi\bpsi_-\bpa\pa\psi_-\bigg)(\beps_-\psi_+)	\notag\\
&-&(\beps_-\bpa\psi_+) [\bpa\phi(\pa\phi)^2 +\bpa\phi(\bpsi_+\pa\psi_+)]	\bigg]
\eea
Integrating a $\bpa$ by parts in the middle term in the first line, we obtain:
\bea
\d_0 L_1 &=& 2\bigg[(\bpa\phi\pa\phi)(\beps_-\psi_+)\pa\bpa\phi - (\beps_-\psi_+)\pa\phi\bpa\bpsi_-\pa\psi_-	\notag\\
&-&(\beps_-\bpa\psi_+) [\bpa\phi(\pa\phi)^2 +\bpa\phi(\bpsi_+\pa\psi_+) + \pa\phi(\bpsi_-\pa\psi_-)]	\bigg]
\eea
Note that there is an ambiguity in the final term, as it is proportional to two different equations of motion. We are free to pick our poison of preference, and will regard this term as a modification to the $\psi_+$ transformations without loss of generality. We subsequently report:
	\begin{align}
		\delta_1 \phi &=+  \pa\phi \bpa\phi  \beps\psi	\nn\\
		\delta_1\bpsi_+ &= +  (\pa\phi)^2 \bpa\phi \beps_- + \big(S_{+,z}\bpa\phi +S_{-,z}\pa\phi\big)\beps_- + (\beps_+\psi_-)\bpa\phi\pa\bpsi_+\lab{del1phipsi}	\\
		\delta_1\bpsi_- &= +  (\bpa\phi)^2 \pa\phi \beps_+ + \big(S_{-,\bz}\pa\phi +S_{+,\bz}\bpa\phi\big)\beps_+ + (\beps_-\psi_+)\pa\phi\bpa\bpsi_-	\nn
	\end{align}
We can write these transformation rules in the following covariant form:
	\begin{align}
		\delta\phi &= \left[1+\lambda \pa\phi\bar\pa\phi\right]\bar\eps\psi	+ \mathcal{O}(\lambda^2)\\
		\delta\psi &= \left[1 - \lambda\pa\phi\bar\pa\phi\right]\slashed{\pa}\phi\,\eps - \lambda \left(\bar\psi\slashed{\pa}\phi\pa_\m\psi\right)\gamma^\m\eps + \lambda\left(\bar\eps\slashed{\pa}\phi\gamma_\m\psi\right)\slashed{\pa}\gamma^\mu\psi + \mathcal{O}(\lambda^2)
	\end{align} 
	
\subsection{$ \d_0 L_2 +\d_1L_1=0$ }

In this section we outline the calculation which establishes supersymmetry at order $\l^2$.  We simplify expressions by simply dropping the EoM terms that we encounter. They can be cancelled by  second order variations as done in the previous section, but we do not compute these variations.  Again we treat only $\beps_-$ terms explicity. 

First we calculate linear variations (in $\psi$) of the two terms in $L_2$ in \reef{lamseries} by a process that also captures the cubic terms.  The $\beps_-\psi_+$ variations of the first term in $L_2$ come from the bilinear $(\bpsi_+\bpa\psi_+)$.
Using \reef{Ttrf} we find:
\be\lab{dL2a}
\d[(\pa\phi\bpa\phi)^2(\bpsi_+\bpa\psi_++ \bpsi_-\pa\psi_-)] =2(\bpa\phi\pa\phi)^2[-\bpa(\beps_-\pa\phi\psi_+) +2\beps_-\bpa\pa\phi\,\psi_+]. 
\ee
The final $[...]$ contains only EoM terms, so we simply drop them. In the second term of $L_2= 2(\bpa\phi\pa\phi)T_{++}T_{--}$, only the variations of the first and second factors contain possible on-shell $\beps_-$ terms. We write ( after partial integration of $\pa$ in the second term)
\bea
2\bpa\phi(\beps_-\pa\psi_+)T_{++}T_{--} -2\bpa\phi\pa\pa\phi(\beps_-\pa\phi\psi_+)T_{--}
\eea
In the second term we write $2\pa\pa\phi\pa\phi = \pa(\pa\phi)^2$, and integrate $\pa$ by parts again, obtaining 
\be\lab{d0L2}
\d_0L_2=2 \bpa\phi(\beps_-\pa\psi_+)T_{++}T_{--}+ \bpa\phi(\pa\phi)^2(\beps_-\pa\psi_+)T_{--}.
\ee
All cubic terms are included in this expression (cubic terms from the $\d\phi $ variation of \reef{dL2a} vanish on-shell since $T_{+-},~T_{-+}$ vanish).

Now is the time for all good men to compute $\d_1L_1$.  Let's concentrate first on the effect of the following terms from \reef{del1phipsi}: 
\be
\d_1\phi =\bpa\phi\pa\phi\beps_-\psi_+\qquad\qquad \d_1\bpsi_+ = \bpa\phi T_{++} \beps_-.
\ee
Note that the bilinear $T_{-+} = S_{-,z}$ vanishes on-shell and can be ignored. We start with  
\be
\d_1 L_{1b} \equiv =-\d_1(T_{++}) T_{--} - T_{++}\d_1T_{--}\,.
\ee
The last variation contains only EoM terms after integrating a $\bpa(\bpa\phi)^2$ by parts, so we continue with 
\bea\lab{d1L1b}
\d_1 L_{1b} &=&-\bpa\phi [2 T_{++}+ (\pa\phi)^2](\beps_-\pa\psi_+) T_{--}
\eea
Two partial integrations were done to obtain this form, and EoM terms were dropped as usual.  We see that \reef{d0L2} is canceled in entirety. Note that all linear terms in $\psi$ have now canceled at order $\l^2$, and some cubic terms have come along for the ride and canceled as well.

So far we have ignored the last term in $\d\bpsi$ in \reef{del1phipsi}.  This term is quadratic in $\psi$ and potentially leads to cubic contributions in $\d_1L_1$. One can see explicitly that these vanish because of the Grassmann property. Quintic contributions meet the same fate. Our job is done; the action \reef{Lsup2} satisfies (1,1)  supersymmetry at least though order $\l^2.$

\subsection{The SUSY algebra at  order $\l$}


As a further check on our work, we explore the SUSY algebra to order $\l$.  We will show that the conventional form of the algebra,  namely 
\begin{align}
		[\delta^1,\delta^2] = -2\beps^1\gamma^\mu\eps^2\pa_\mu ,\label{eqn:SUSYAlgebra}
	\end{align}
holds on all fields of the theory. Order $\l$ terms occur in intermediate stages of the computation, but they cancel on-shell, specifically when fields satisfy their equations of motion through order $\l$.
We start with the scalar $\phi$ which is the simpler case:
\begin{align}
	\delta^1\delta^2\phi &= (1+\l\pa\phi\bar\pa\phi)[\beps_+^2\eps_+^1\bpa\phi + \beps_-^2\eps_-^1\pa\phi] + \l[\beps_+^1\bpa\psi_-\pa\phi + \beps_-^1\pa\psi_+\bpa\phi][\beps_+^2\psi_- + \beps_-^2\psi_+]	\notag\\
		&	-\l\beps_+^2[(\bpa\phi)^2\pa\phi\eps_+^1 + S_{-,\bz}\pa\phi\eps_+^1 - (\beps_-^1\psi_+)(\pa\phi)(\bpa\psi_-)]	\notag\\
		&	-\l\beps_-^2[(\pa\phi)^2\bpa\phi\eps_-^1 + S_{+,z}\bpa\phi\eps_-^1 - (\beps_+^1\psi_-)(\bpa\phi)(\pa\psi_+)]+\co( \l^2).	
\end{align}
 All purely bosonic terms of order $\l$ cancel. 
We now reorder the remaining terms, minding minus signs from Grassmann variable exchanges and Majorana flips, to obtain the following form:
	\begin{align}
		\delta^1\delta^2 \phi &= -\beps^1\gamma^\mu\eps^2 \pa_\mu\phi	 + \l\bigg[ -(\beps_+^1\eps_+^2 + \beps_+^2\eps_+^1)S_{-,\bz}\pa\phi + (\beps_+^1\beps_-^2 + \beps_+^2\beps_-^1)\psi_+\bpa\psi_-\pa\phi	+ \text{L.c.}\bigg]
	\end{align}
The order-$\l$ terms are symmetric under interchange of $1 \lra 2$ and therefore cancel in the commutator $[\delta^1,\delta^2]$.
This leaves us with Eq.~(\ref{eqn:SUSYAlgebra}) as desired. 

We now attack the $\psi$ case. We focus on $\psi_+$, since this determines the form of the commutator acting on $\psi_-$ by Lorentz conjugation. Before we begin, we should specify the equations of motion for $\psi_+$ and $\psi_-$ from $L_0+\l L_1$. In the following expressions, we have made a helpful simplification: all order-$\l$ terms proportional to lowest-order EoMs can be exchanged for terms which are $\co(\l^2)$ on equations of motion. We find:
	\begin{align}
		\bpa\psi_+ &= \l T_{--} \pa\psi_+  + \mathcal{O}(\l^2)	\\
		\pa\psi_- &= \l T_{++} \bpa\psi_- + \mathcal{O}(\l^2)
	\end{align}
We will need this information shortly.

Now we study the action of $\delta^1\delta^2$ on the fermion. 
After implementing obvious cancelations, we find
	\begin{align}
		\delta^1\delta^2\psi_+ &= \beps_-^1\pa\psi_+\eps_-^2 - \beps_+^1\beps_-^2(\overbrace{\pa\psi_- - \l\bpa\psi_-T_{++}}^{\mathcal{O}(\l^1)\text{ EoM}}) + \l\pa\pa\phi\bpa\phi(\beps^1_+\psi_-)\eps_-^2	\notag\\[0.25em]
	&\quad\quad +\l\beps_+^2\eps_+^1\underbrace{T_{--}\pa\psi_+}_{\text{EoM}} - \l\beps_+^2\eps_-^1\psi_-\bpa\phi\pa\pa\phi
	\end{align}
The last line arises  from the action of $\d^1$  on the $\psi_+\psi_-$ mixing term of \reef{del1phipsi}. We cancel the order-$\l$ equation of motion in the first line and replace the order-$\l$ EoM-like term in the second line with a $\bpa\psi_+$:
	\begin{align}
		\delta^1\delta^2\psi_+ &= -\beps_-^1\eps_-^2\pa\psi_+ +\beps_+^2\eps_+^1\bpa\psi_+ - \l(\beps_+^2\eps^1+\beps^1_+\eps_-^2)\psi_-\bpa\phi\pa\pa\phi 
	\end{align}
In the commutator, the last term cancels. For the remaining terms, we apply a Majorana flip in the second term (giving an overall minus sign) and take the commutator. We are left with exactly Eq.~(\ref{eqn:SUSYAlgebra}).

Intriguingly, we used  the order-$\l$ equations of motion in two different ways. One involved a full cancelation of terms at order $\l$, and the other required addition and subtraction  of a zero-order term to obtain the required structure in the algebra. 

\section{Discussion}
In this paper, we have outlined a general method to construct $T \bar{T}$-deformed classical actions in a completely algebraic manner.
It turns out that imposing the classical gravitational equations of motion of topological gravity and re-expressing the undeformed action in the new target space coordinates introduced in the gravity path integral exactly solves the defining differential equation \reef{floweq} of Zamolodchikov.
This derivation is suggestive of the general picture of the $T \bar{T}$ deformation as a `movie' of the original theory \cite{dubovsky} --- it can be thought of as the original theory seen from the `wrong' manifold.

The algebraic procedure that derives from these equations of motion, first emphasised in \cite{Conti}, can be solved in a straightforward manner.
More concretely, all of the results in Sec.~3 were generated by changing one line in the same Mathematica file and letting it run.

Further, we used the method above to construct the deformed Lagrangian for a set of $n$ scalar multiplets of $(1,1)$ supersymmetry. For one multiplet, we showed that this Lagrangian is (off-shell) supersymmetric to order $\l^2$.   Specifically supersymmetry holds with modified transformation rules which we wrote explicitly to order $\l$. We believe that a continuation of our approach would yield off-shell invariance at all orders in perturbation theory, and we hope to address this claim in future work.
The next step in the program of explicit $T\bar  T$ constructions appears to be the supersymmetric non-linear $\s$-model.

Another potentially interesting direction is to find a coordinate transformation in superspace that implements the manifestly supersymmetric deformation of the transformations~\cite{Baggio,Sethi}.
So far, we have not found a working set of defining equations for this coordinate transformation.

\section{Acknowledgements}
We thank Louise Anderson, Victor Gorbenko, Edward Mazenc and Vasudev Shyam for illuminating discussions.
We especially thank Eva Silverstein for unfailing interest and guidance. EAC thanks John Cardy, Brandon Rayhaun, and Jonathan Sorce for insightful comments.

JAD would like to thank the Stanford Institute for Theoretical Physics for its hospitality during the first stage of this project. The work of EAC is supported by the US NSF Graduate Research Fellowship under Grant DGE-1656518. The work of JAD is supported by CONICET and by a Fulbright - Bunge y Born Fellowship.
The research of DZF is partially supported by US NSF grant Phy-1620045.

\bibliographystyle{JHEP}
\bibliography{refs}

\end{document}